\documentclass[draft]{agujournal2019}
\usepackage{url} 
\usepackage{lineno}
\usepackage[finalnew]{trackchanges} 
\usepackage{soul}
\draftfalse

\journalname{Geophysical Research Letters}

\begin{document}

%
%

\title{Ganymede MHD Model: Magnetospheric Context for Juno's PJ34 Flyby}

%
%




\authors{Stefan Duling\affil{1}, Joachim Saur\affil{1}, George Clark\affil{2}, Frederic Allegrini\affil{3}, Thomas Greathouse\affil{3}, Randy Gladstone\affil{3,4}, William Kurth\affil{5}, John E. P. Connerney\affil{6,7}, Fran Bagenal\affil{8}, Ali H. Sulaiman\affil{9}}

\affiliation{1}{Institute of Geophysics and Meteorology, University of Cologne, Cologne, Germany}
\affiliation{2}{The Johns Hopkins University Applied Physics Laboratory, Laurel, Maryland, USA}
\affiliation{3}{Southwest Research Institute, San Antonio, Texas, USA}
\affiliation{4}{University of Texas at San Antonio, San Antonio, Texas, USA}
\affiliation{5}{Department of Physics and Astronomy, University of Iowa, Iowa City, Iowa, USA}
\affiliation{6}{NASA Goddard Space Flight Center, Greenbelt, Maryland, USA}
\affiliation{7}{Space Research Corporation, Annapolis, Maryland, USA}
\affiliation{8}{Laboratory for Atmospheric and Space Physics, University of Colorado, Boulder, Colorado, USA}
\affiliation{9}{School of Physics and Astronomy, Minnesota Institute for Astrophysics, University of Minnesota, Minneapolis, Minnesota, USA}
 




\correspondingauthor{Stefan Duling}{stefan.duling@uni-koeln.de, https://orcid.org/0000-0001-7220-9610}




\begin{keypoints}
\item Our MHD model illustrates the state of Ganymede's magnetosphere during Juno's flyby and locates its trajectory outside closed field lines.
\item The location of the open-closed-field line-boundary is predicted and matches the poleward edges of the aurora as observed by Juno.
\item We investigate model uncertainties caused by incomplete knowledge of upstream conditions and other parameters.

\end{keypoints}

%
%

%
%

\begin{abstract}
On June 7th, 2021 the Juno spacecraft visited Ganymede and provided the first in situ observations since Galileo's last flyby in 2000. The measurements obtained along a one-dimensional trajectory can be brought into global context with the help of three-dimensional magnetospheric models. Here we apply the magnetohydrodynamic model of \citeA{Duling2014} to conditions during the Juno flyby. In addition to the global distribution of plasma variables we provide mapping of Juno's position along magnetic field lines, Juno's distance from closed field lines and detailed information about the magnetic field's topology. We find that Juno did not enter the closed field line region and that the boundary between open and closed field lines on the surface matches the poleward edges of the observed auroral ovals. To estimate the sensitivity of the model results, we carry out a parameter study with different upstream plasma conditions and other model parameters.
\end{abstract}

\section*{Plain Language Summary}
In June 2021 the Juno spacecraft flew close to Ganymede, the largest moon of Jupiter, and explored its magnetic and plasma environment. Ganymede's own magnetic field forms a magnetosphere, which is embedded in Jupiter's large-scale magnetosphere, and which is unique in the solar system. The vicinity of Ganymede is separated into regions that differ in whether the magnetic field lines connect to Ganymede's surface at both or one end or not at all. These regions are deformed by the plasma flow and determine the state of the plasma and the location of Ganymede's aurora. We perform simulations of the plasma flow and interaction to reveal the three-dimensional structure of Ganymede's magnetosphere during the flyby of Juno. The model provides the three-dimensional state of the plasma and magnetic field, predicted locations of the aurora and the geometrical magnetic context for Juno's trajectory. These results are helpful for the interpretation of the in situ and remote sensing obtained during the flyby. We find that Juno did not cross the region with field lines that connect to Ganymede's surface at both ends. Considering possible values for unknown model parameters, we also estimate the uncertainty of the model results.

\section{Introduction} 
As the largest moon in the solar system, Ganymede not only resides inside Jupiter's huge magnetosphere but also possesses an intrinsic dynamo magnetic field \cite{Kivelson1996}. The co-rotating Jovian plasma overtakes Ganymede in its orbit with sub-alfv\'enic velocity and drives an interaction that is unique in the solar system. The internal field acts as an obstacle for the incoming plasma flow, generating plasma waves, Alfv\'en wings and electric currents along the magnetopause \cite{Gurnett1996,Frank1997,Williams1997}. The incoming Jovian magnetic field reconnects at the boundary of a donut-shaped equatorial volume of closed field lines that are defined by both ends connecting to Ganymede's surface \cite{Kivelson1997}. The open field lines in the polar regions connect to Jupiter at the other end and define the extent of Ganymede's magnetosphere. Near the open-closed-field line-boundary (OCFB) observations by Hubble Space Telescope (HST) revealed the presence of two auroral ovals within Ganymede's atmosphere \cite{Hall1998, Feldman2000}.

On June 7th, 2021 Juno approached Ganymede from the downstream side and crossed the magnetospheric tail for the first time. Juno encountered Ganymede with a minimum distance of 1046km ($\sim$0.4 radii) on a trajectory heading northwards and towards Jupiter, leaving the interaction system at its flank \cite{Hansen2022}.

For analyzing and interpreting the measurements obtained by Juno \cite{Allegrini2022,Clark2022,Kurth2022} it is important to study which part of its trajectory is geometrically related to the various regions of Ganymede's magnetosphere. Juno's measurements could not uniquely conclude whether Juno crossed the closed field line region. For example, JEDI found double loss cones for $>$30keV electrons \cite{Clark2022} while JADE found only single loss cones \cite{Allegrini2022}. To find the location where detected particles can interact with Ganymede's atmosphere or surface, i.e. Juno's magnetic footprint, the necessary field line tracing requires a model for the magnetic field. Furthermore, Juno's UVS instrument provided auroral images at unprecedented resolution \cite{Greathouse2022}. Electron acceleration processes driving Ganymede's aurora are not fully understood, however, from analysis of poorly resolved HST observations it was argued that the aurora occurs near the OCFB \cite{McGrath2013}. To substantiate this previous finding a comparison of the Juno UVS observations with the modeled magnetic topology is of considerable interest. The aim of this work is thus to provide field and mapping properties during the flyby and illustrate the three-dimensional context of Juno's measurements (Section 3). We further carry out a model sensitivity study on uncertain upstream conditions and other model parameters to estimate their impact and the uncertainty of our results (Section 4).

\begin{figure}
\noindent\includegraphics[width=\textwidth]{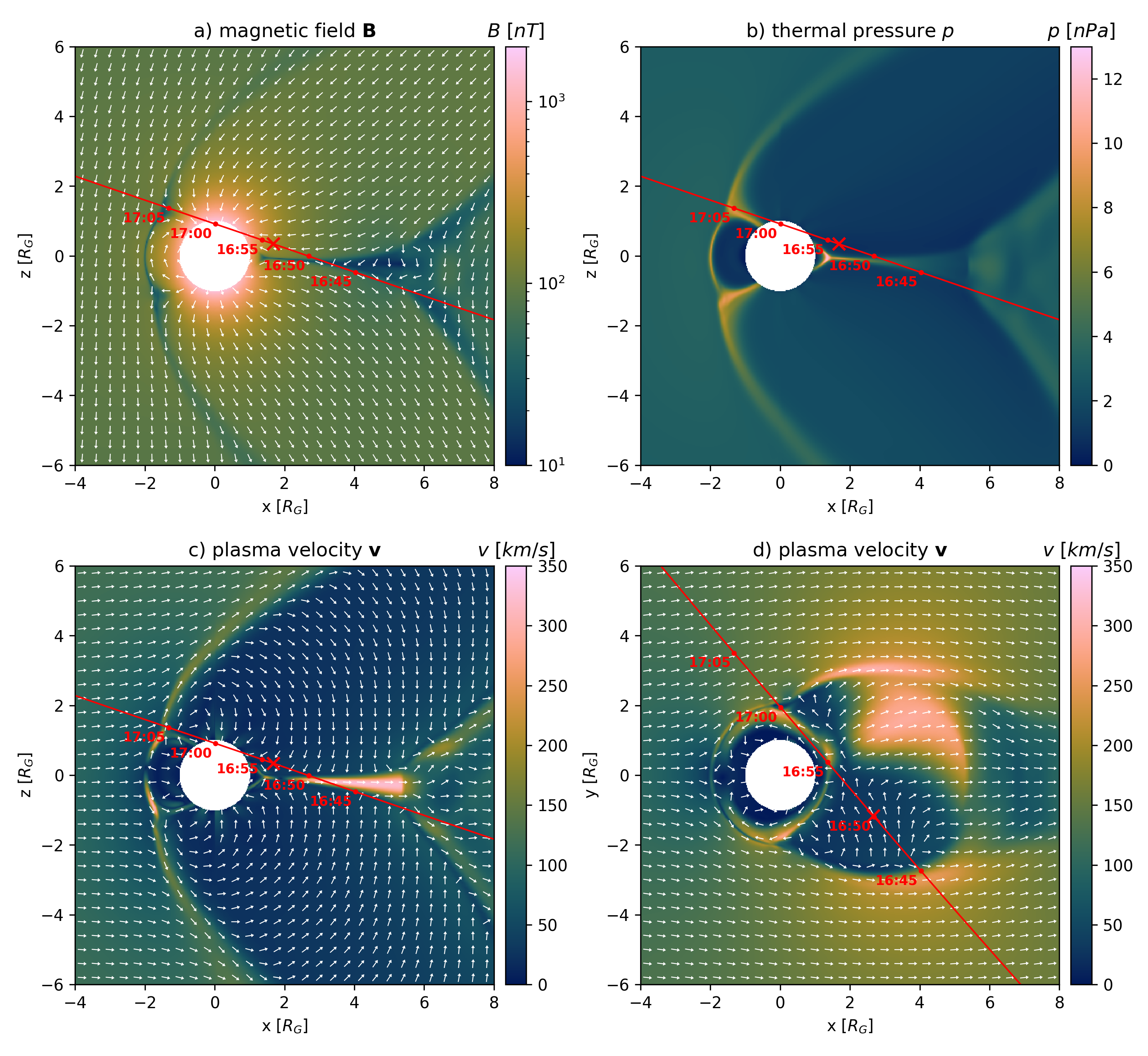}
\caption{Model variables for Juno's flyby, plasma flow from left to right. $y$=0 plane: a) Magnetic field $\mathbf{B}$, b) thermal pressure $p$, c) velocity $v$. Equatorial plane, $z$=0: d) velocity $\mathbf{v}$. The red crosses indicate Juno's crossing through these planes and the red lines the projected trajectory. The white arrows show the projected direction of $\mathbf{B}$ and $\mathbf{v}$ respectively. Figures S3-S4 show planes with minimized trajectory projection.}
\label{2dplots}
\end{figure}

\section{Model} 
\label{chmodel}

We describe Ganymede's space environment by adopting a magnetohydrodynamic (MHD) model based on \citeA{Duling2014}, which describes a steady state solution for a fixed position in Jupiter's magnetosphere. In our single-fluid approach the plasma interaction is described by the plasma mass density $\rho$, plasma bulk velocity $\mathbf{v}$, total thermal pressure $p$ and the magnetic field $\mathbf{B}$. For these variables appropriate boundary conditions are applied at Ganymede's surface and at a distance of 70 Ganymede radii ($R_G=2631$ km). Our model includes simplified elastic collisions with an O$_2$ atmosphere, ionization processes and recombination. Ganymede's intrinsic magnetic field is described by dipole Gauss coefficients $g_1^0=-716.8$ nT, $g_1^1=49.3$ nT, $h_1^1=22.2$ nT \cite{Kivelson2002}. During Juno's visit Ganymede was near the center of the current sheet (302$^{\circ}$W System-III, -2$^{\circ}$ magnetic latitude) where the induction response of an expected ocean \cite{Saur2015} is close to minimum. In our model the induced field has a maximum surface strength of 15.6 nT. The upstream plasma conditions are adjusted to the flyby situation as listed in Table \ref{table}. They characterize the interaction to be sub-Alfv\'enic with an Alfv\'en Mach number of $M_A=$0.8 and a plasma beta of 1.1. 

While we utilized the ZEUS-MP code \cite{Hayes2006} in \citeA{Duling2014} we now present results obtained with the PLUTO code \cite{Mignone2007}. Simulating the identical physical model with both independent solvers produces similar results (S4), suggesting additional reliability. It also enables us to estimate the uncertainties due to different numerical solvers, never done before in Ganymede's case. A detailed description of our model (S1), a discussion of the uncertainty of upstream conditions and model parameters (S2) and the numerical implementation (S3) is attached in the Supplementary Information. We use the GPhiO coordinates, where the primary direction $z$ is parallel to Jupiter's rotation axis, the secondary direction $y$ is pointing towards Jupiter and $x$ completes the right-handed system in direction of plasma flow.

\section{Results} 
\label{chresults}

\begin{figure}
\noindent\includegraphics[width=\textwidth]{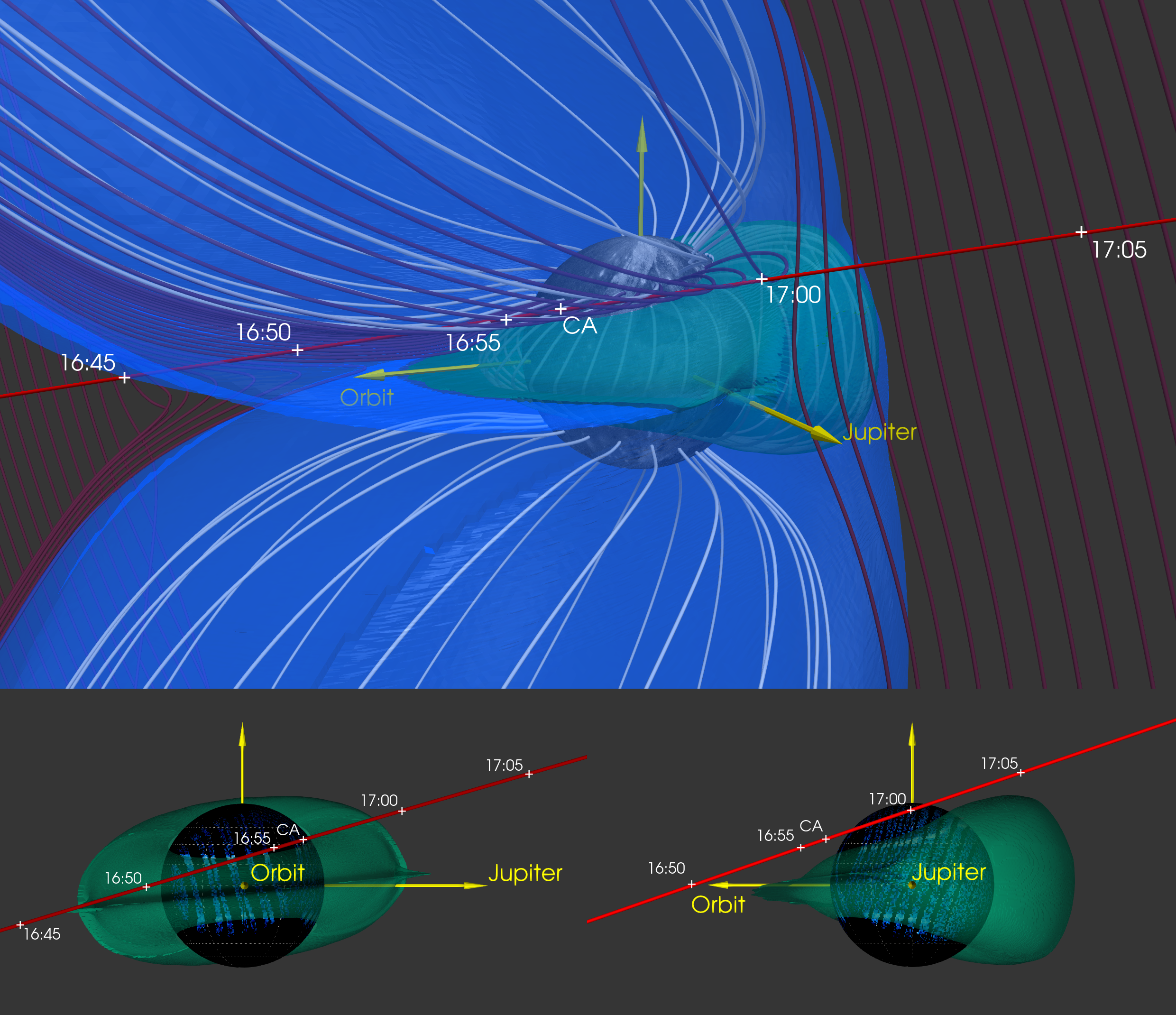}
\caption{Juno's trajectory (red) in relation to the modeled magnetosphere during the flyby of Ganymede. The timestamps in UTC indicate the position of Juno. In the upper panel the lines show selected magnetic field lines connected to Ganymede's surface (white) and Juno's trajectory (dark). The green surface represents the outer boundary of closed field lines, the blue surface represents the outer boundary of open field lines that connect to Ganymede at one end. The bottom panel additionally shows observed 130.4 and 135.6 nm oxygen emissions from the aurora in blue \cite{Greathouse2022}.}
\label{3dplot}
\end{figure}

For the time of closest approach (CA) the Jovian background magnetic field was inclined by $\sim$20° to Ganymede's spin and by $\sim$15° to Ganymede's dipole axis, leading to a sub-alfv\'enic interaction that is roughly symmetric to the $y=0$ plane. Ganymede's magnetosphere is characterized by northern and southern Alfv\'en wings, both bent in the orbital direction. In Figure \ref{2dplots} they can be identified by a tilted magnetic field and lowered plasma velocity and pressure. The modeled angle ($xz$-plane projection) between the northern wing and the $z$ axis of $\sim$46° matches the theoretical value of 46.5° based on the theory of \citeA{Neubauer1980}. Inside the Alfv\'en wings the plasma velocity is reduced below 50 km/s. The convection through the wings over the poles is slowed and takes about 10 minutes for a distance of 10 $R_G$. The interaction expands the volume characterized by closed field lines on the upstream side in $z$ direction while it is strongly compressed on the downstream side. This area has a thermal pressure below 1 nPa in Figure \ref{2dplots}b. The diameter of Ganymede's magnetosphere is about 4 $R_G$ in the equatorial plane as indicated by the reduced and reversed velocity in Figure \ref{2dplots}d. On the downstream side the reduced velocity also indicates a stretched magnetospheric tail with more than 10 $R_G$ length that was crossed by Juno at the location of the red cross.

\subsection{Magnetic Topology}
\label{chresultstopology}

\begin{figure}
\noindent\includegraphics[width=\textwidth]{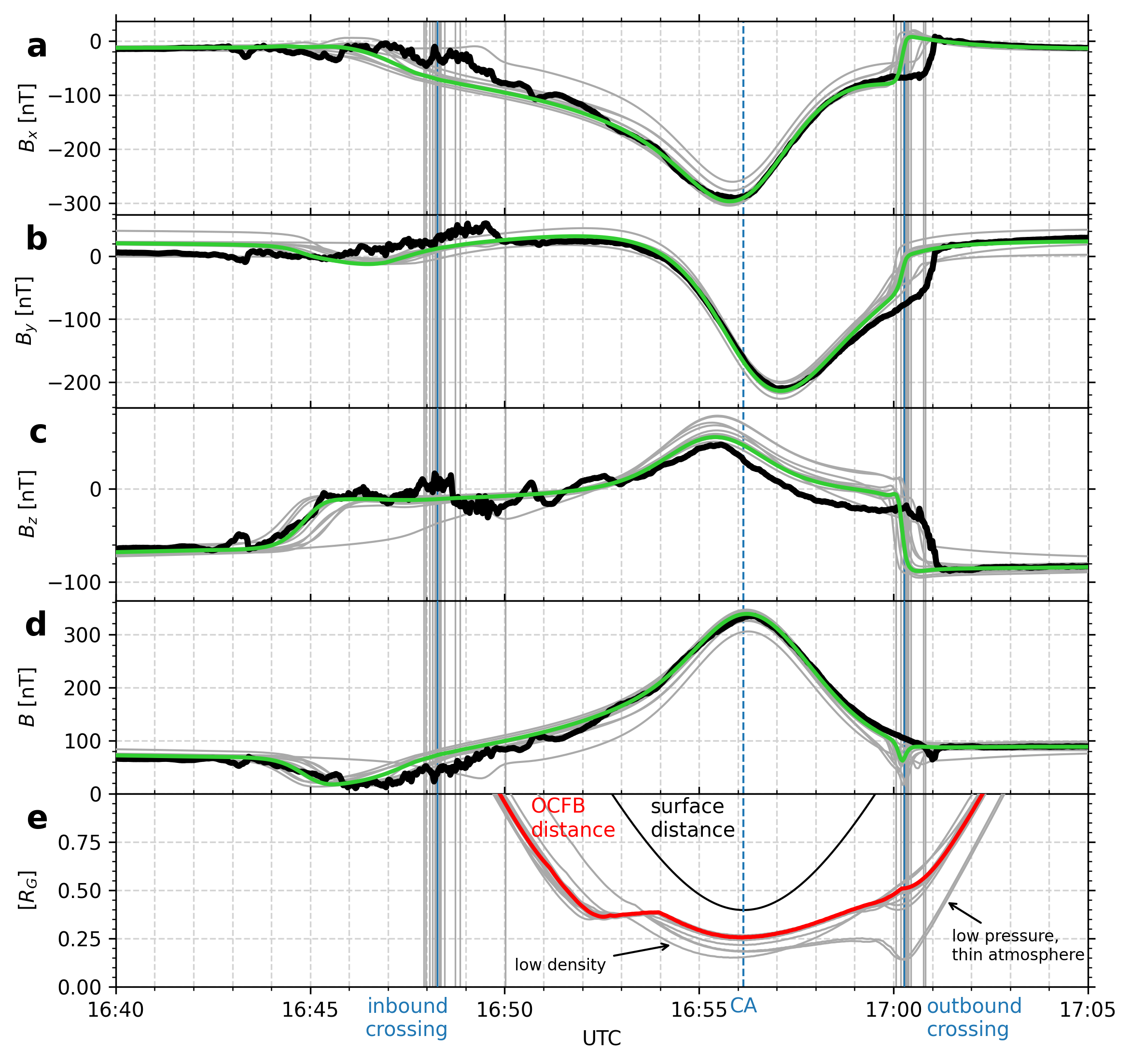}
\caption{Modeled (green) versus measured (black) magnetic field along Juno's trajectory (panels a-d, GPhiO). Panel e shows Juno's distance from Ganymede's surface (black) and the OCFB (red) in $R_G$. The blue vertical lines represent the modeled inbound and outbound magnetopause crossings. The gray lines indicate model uncertainty from uncertain upstream conditions (Section \ref{chrobustness}).}
\label{trajplot}
\end{figure}

In Figure \ref{3dplot} and Movie S1, we display the modeled magnetic field topology together with Juno's trajectory (red) in 3D. The volume of open field lines is represented by the blue surface and was crossed by Juno. In our model the crossings occurred inbound at 16:48:16 on the tail side and at 17:00:16 outbound at the northern Jupiter-facing side. We do not see Juno on closed field lines at any time. The height of the closed field line region, green in Figure \ref{3dplot}, increases in upstream direction. Juno's trajectory is located slightly above this boundary and inclined by a similar angle. Therefore the closest distance between Juno and closed field lines was relatively constant below 0.4 $R_G$ for about 7 minutes, with a minimum of $\sim$0.26 $R_G$ at the time of CA (Figure \ref{trajplot}).

\begin{figure}
\noindent\includegraphics[width=\textwidth]{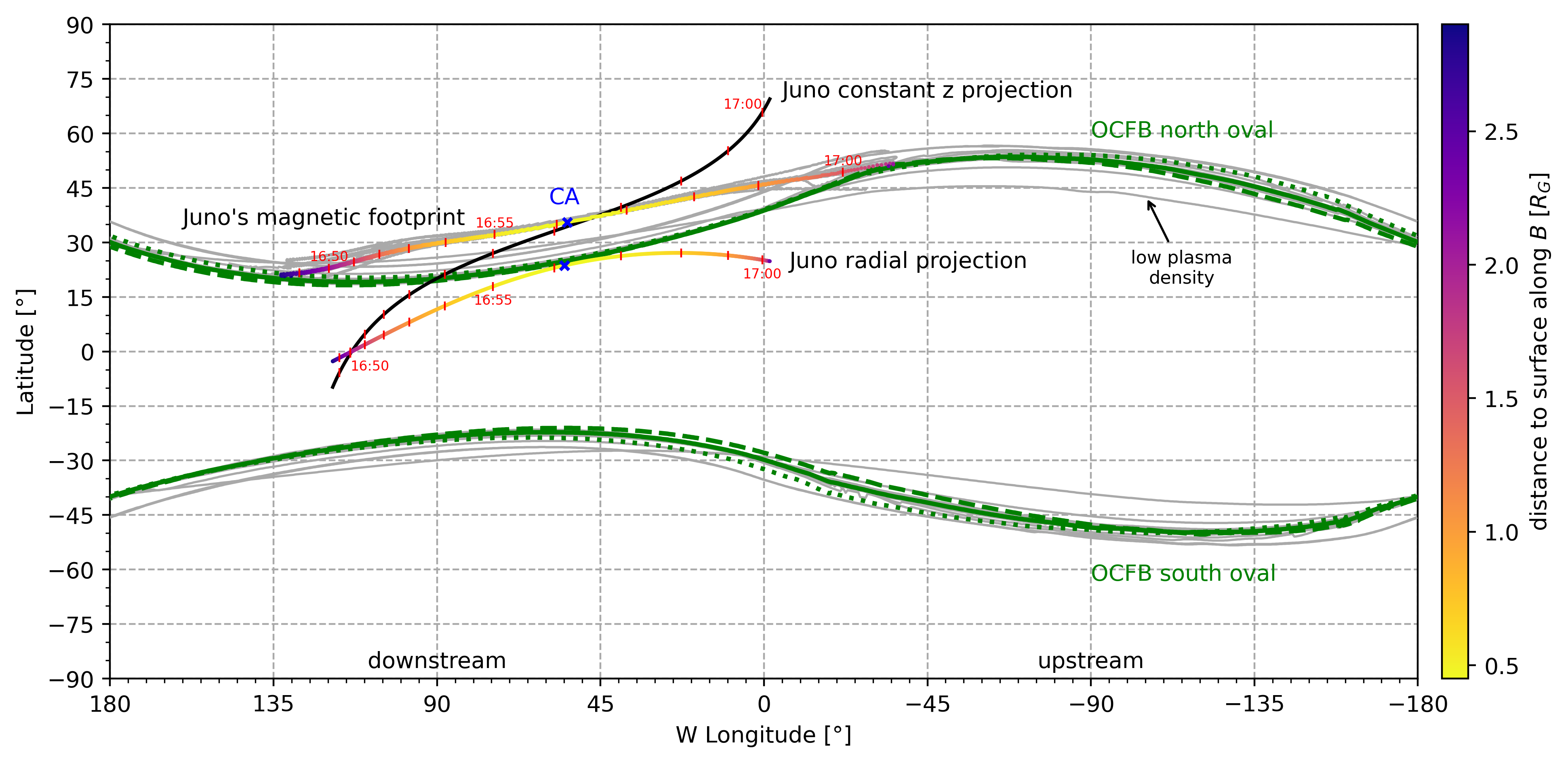}
\caption{Surface map of Ganymede with 0° longitude pointing towards Jupiter ($+y$ GPhiO). The modeled OCFB from our default model is shown as solid green lines. The dotted (dashed) lines show its location based on modelling with the measured background field before (after) the flyby. \citeA{Greathouse2022} show the coincidence of the OCFB and the observed aurora. Juno's position, while inside the magnetosphere, is projected in radial direction and shown as the lower multicolored line, the same with constant $z$ as black line. The upper multicolored line shows the location where field lines end that are connected to Juno, namely Juno's magnetic footprint. Color coded is the distance along those field lines. The gray lines indicate model uncertainty from uncertain upstream conditions (Section 4).}
\label{auroraplot}
\end{figure}

The two solid green lines in Figure \ref{auroraplot} show the location of the OCFB on Ganymede's surface, calculated by field line tracing. The plasma flow generates magnetic stresses which push the OCFB pole-wards on the upstream side and press them together on the downstream side. Here the averaged latitude (between 45° and 135° W longitude) is at 21.2° (north) and -24.4° (south), respectively. \citeA{Greathouse2022} compare the OCFB location with Ganymede's aurora observed by Juno, summarized in Section \ref{chsummary}. Figure \ref{auroraplot} also shows results from alternative simulations with the background field measured approximately 30 minutes before (dotted) and after (dashed) the flyby. As consequence of the field rotation the OCFB lines appear to migrate in opposite directions, west for the northern and east for the southern hemisphere. This is also identifiable by the longitudinal migration of the latitudinal minimums and maximums (before$\mid$CA$\mid$after): 108°$\mid$111°$\mid$113° and -88°$\mid$-70°$\mid$-67° (north), 62°$\mid$60°$\mid$53° and -107°$\mid$-117°$\mid$-121° (south).

The lower multicolored line in Figure \ref{auroraplot} shows Juno's radially projected trajectory inside the magnetosphere, its endpoints refer to the magnetopause crossings. The crossings also correspond to the blue vertical lines in Figure \ref{trajplot} and the punctures of the blue surface in Figure \ref{3dplot}. Tracing the field lines from Juno's position to the surface yields its magnetic footprint, as shown as upper multicolored line in Figure \ref{auroraplot}. Since the colors indicate the lengths of the field lines between Juno and the surface, the footprint location associated to a fixed position of the spacecraft can be identified by a shared color. Juno's footprint is modeled to be up to 11° and on average 7° degree north of the OCFB as modeled with the estimated background field during CA. During approach to CA Juno's mapped position on the surface was nearly on the same meridian as Juno itself. After CA the field lines become more bent in longitudinal direction (Figure \ref{3dplot}) resulting in an eastern shift of Juno's footprint. Juno's footprint touches the OCFB at both ends. While this is counter intuitive at first glance, it is a direct consequence of the magnetic topology. Every magnetopause crossing, although possibly far away from closed field lines, touches an open field line that ends at the OCFB at Ganymede's surface. This convergence of field lines brings the footprints on the surface closer to the OCFB than Juno's position itself.

\subsection{Comparison with Magnetometer Measurements}
\label{chmodeltrajectory}

In Figure \ref{trajplot} we compare our modeled magnetic field with Juno's magnetometer (MAG) measurements \cite{Connerney2017}. The blue vertical lines represent the modeled times when Juno entered and left the open field line region, namely the inbound and outbound magnetopause crossings. Although short-term fluctuations are not covered by our model, the overall structure is reproduced very well. The field rotations have a consistent shape and even the rotation in the wake region (16:45) is predicted at the correct time. The latter demonstrates that the increased diameter of the tail structure (Figure \ref{2dplots}d) is consistent with the observations. This feature is sensitive to the spatial resolution (see Figure S1). During the actual inbound magnetopause crossing, both the measurements and our model do not indicate a rotation.

We identify two noticeable deviations. (1) The model features a clear outbound crossing but it is located slightly too far inwards and occurs $\sim$40 s too early. We analyze the impact of uncertain upstream conditions on this in Section \ref{chrobustness}. (2) In the closer vicinity of Ganymede $B_z$ is slightly overestimated by 10 to 20 nT. We found that this deviation is sensitive to the numerical resolution in latitudinal direction, which affects the compression of the closed field line region on the downstream side. We interpret this that a high resolution is required to resolve the strong magnetic stresses at lower latitudes. Our latitudinal resolution is $\sim$0.75°. We expect the $B_z$ deviation might reduce further if an even higher resolution would be feasible.

\section{Model Sensitivity} 
\label{chrobustness}

For the interpretation of Juno's measurements a model can play an important role. In contrast to measurements, however, it is complex to apply a quantitative error analysis to assess the uncertainty of our quantitative results. Model errors originate from (1) model assumptions, (2) uncertain parameters (this section) and (3) the numerics (Supplementary Information S4).

To investigate error source (2) we carry out a parameter study by varying single parameters to their individual realistic minimum and maximum values as listed in Table 1. This also helps to estimate the model sensitivity on each parameter. The upstream conditions during Juno's flyby are not completely available from direct measurements alone and therefore contain uncertainties of different magnitude, as described in detail in the Supplementary Information (S2). The parameter study also includes uncertainties of the primary dipole moment $g_1^0$ ($\pm$ 2\%) and our parametrizations of the atmospheric density and ionization rate, assuming uncertainties each by a factor of 5.

\begin{table}[!ht]
\caption{Variations of model parameters and upstream conditions and their effect on presented model results. Columns 3-6 specify the averaged latitude of the northern and southern open closed field line boundary (OCFB) on Ganymede's surface on the upstream (-45° to -135°W) and downstream (45° to 135°W) side. Column 7 lists Juno's closest distance to closed field lines (CF) and columns 8-9 the UTC times of its inbound and outbound magnetopause crossings, respectively. Column 10 lists the RMS between measured and modeled magnetic field between 16:50 and 16:59.}
\label{table}
    \centering
    \resizebox{\textwidth}{!}{%
    \begin{tabular}{ l l l l l l l l l r}
        & & \multicolumn{2}{c}{OCFB down}& \multicolumn{2}{c}{OCFB up} & CF & \multicolumn{2}{c}{magnetopause crossing} & RMS\\
        parameter & value & N [°] & S [°] & N [°] & S [°] & [$R_G$] & inbound & outbound & [nT] \\ 
        \hline
        default         & - $^a$             & 21.2 & -24.4 &  51.5 & -47.4 & 0.26 & 16:48:16 & 17:00:16 &  9.3\\
        $B_0$ before CA & (-16,3,-70) nT $^b$& 22.1 & -25.4 &  52.7 & -48.4 & 0.25 & 16:48:43 & 17:00:46 & 12.8\\
        $B_0$ after CA & (-14,43,-80) nT $^b$& 20.6 & -23.7 &  50.6 & -46.7 & 0.26 & 16:48:13 & 17:00:03 & 12.0\\
        velocity       & 120 km/s $^c$       & 22.1 & -25.3 &  50.6 & -46.4 & 0.24 & 16:48:27 & 17:00:22 &  9.7\\
        velocity       & 160 km/s $^c$       & 20.8 & -23.9 &  52.7 & -48.8 & 0.26 & 16:48:04 & 17:00:19 & 11.0\\
        density        & 10 amu/cm$^3$ $^d$  & 26.5 & -30.3 &  43.2 & -38.8 & 0.15 & 16:50:01 & 17:00:49 & 27.4\\
        density        & 160 amu/cm$^3$ $^c$ & 20.6 & -23.6 &  53.3 & -49.5 & 0.27 & 16:47:55 & 17:00:17 & 12.6\\
        pressure       & 1 nPa               & 25.3 & -28.4 &  54.5 & -50.4 & 0.14 & 16:48:12 & 17:00:26 & 19.5\\
        pressure       & 5 nPa               & 21.4 & -24.6 &  50.9 & -46.9 & 0.25 & 16:48:51 & 17:00:11 &  9.7\\
        production     & 0.5e-8 /s           & 21.3 & -24.5 &  51.7 & -47.6 & 0.25 & 16:48:21 & 17:00:16 &  9.5\\
        production     & 10e-8 /s            & 21.7 & -24.9 &  51.4 & -47.3 & 0.24 & 16:48:08 & 17:00:18 &  9.8\\
        atmosphere     & 1.6e6 /cm$^3$       & 25.4 & -28.5 &  54.7 & -50.6 & 0.14 & 16:48:19 & 17:00:23 & 19.4\\
        atmosphere     & 40e6 /cm$^3$        & 23.4 & -26.7 &  48.6 & -44.6 & 0.22 & 16:47:58 & 17:00:20 & 13.1\\
        dynamo $g_1^0$ & -2\%                & 21.1 & -24.3 &  51.9 & -47.8 & 0.26 & 16:48:16 & 17:00:16 &  9.3\\
        dynamo $g_1^0$ & +2\%                & 21.7 & -24.8 &  51.9 & -47.9 & 0.25 & 16:48:14 & 17:00:19 & 10.4\\
        \multicolumn9l{$^a$: default values: (-15,24,-75) nT $^b$, 140 km/s $^c$, 100 amu/cm$^3$, 2.8 nPa $^c$, 2.2e.8 /s, 8e6 /cm$^3$}\\
        \multicolumn9l{$^b$: \citeA{Weber2022}}\\
        \multicolumn9l{$^c$: \citeA{Kivelson2022}}\\
        \multicolumn9l{$^d$: \citeA{Bagenal2011}}\\
    \end{tabular}}
\end{table}

In the MHD view the locations of magnetopause and OCFB are determined by equilibriums of forces that depend on the physical parameters of the model. Table \ref{table} summarizes the sensitivities of important model results to different parameter variations that are each displayed as gray lines in Figures \ref{trajplot} and \ref{auroraplot}. A significantly later outbound magnetopause crossing (17:00:49 latest) is modeled if the upstream plasma density is extraordinary low or the measured background field before CA is used. The latter is unlikely to still represent the background field when Juno crossed the magnetopause about 5 minutes after passing CA. With an uncertainty of $\sim$2 minutes the inbound crossing is more sensitive than the outbound crossing ($\sim$45 seconds), as expected from the more dynamic tail where Juno entered Ganymede's magnetosphere.

Our model does not see Juno on closed field lines for any of the considered parameter variations. As Figure \ref{trajplot}e suggests, the sensitivity of the distance to closed field lines can be divided into two parts. Before $\sim$16:59 the uncertainty is quite constant $<$0.15 $R_G$. After $\sim$16:59, around the outbound crossing, when Juno was above the flank of the closed field line region, the uncertainty is larger and especially low plasma pressure and a thinner atmosphere significantly reduce the distance to closed field lines (0.14 $R_G$). Additionally, but near CA, the distance is also clearly reduced if lower plasma density is used (0.15 $R_G$). However, the impact of reduced density and plasma pressure on the physics is different. A lower upstream plasma pressure directly affects the equilibrium of forces at the magnetopause. For unchanged magnetic pressure a reduced plasma pressure thus globally shifts the magnetopause and inflates the total magnetosphere. This results not only in earlier inbound and later outbound crossings but also increases the closed field line region, evolving a secondary minimal distance to Juno's trajectory near Juno's outbound crossing and globally shifting the surface OCFB polewards by 3-4° (Table \ref{table}). In contrast, a lower upstream density reduces the momentum of the plasma and thus reduces the interaction strength \cite{Saur2013}. As consequence the interaction induced upstream/downstream asymmetry of the closed field line region is weaker. The surface OCFB is shifted 5-6° polewards / 8-9° equatorwards on the downstream / upstream side and Juno's trajectory is closer to closed field lines near CA. Varying the upstream velocity shows similar impact, even if less pronounced due to its weaker uncertainty.

In Figure \ref{auroraplot}, the gray lines show the OCFB location on Ganymede's surface from all simulations with parameter variations. On the upstream side the OCFB location is most sensitive to a reduced density. The total uncertainty from all parameter variations is $\sim$12° upstream and $\sim$7° for the remaining longitudes. However, the plasma density and velocity have a stronger impact upstream, while the production rate mainly affects the downstream side.

Table \ref{table} also lists the deviation of the modeled ($\mathbf{B}$) from the measured ($\hat{\mathbf{B}}$) magnetic fields, defined by $RMS= \sqrt{\frac{1}{3N}\sum_i^N\|\mathbf{B}_i-\hat{\mathbf{B}}_i\|^2}$. We emphasize that this alone is not an appropriate method to assess a model's capability to reproduce measurements; e.g. models that reproduce measured field rotations slightly shifted in time might have a higher RMS than models without any rotations at all. Therefore we consider only the interval 16:50-16:59 to exclude the predicted boundary crossings. According to this evaluation we find that the default parameter setup indeed fits the MAG data best and the variations that reduce the distance to closed field lines have a strongly increased deviation.

\section{Discussion and Conclusions} 
\label{chsummary}

We performed MHD simulations of Ganymede's magnetosphere which put Juno's observations into a three-dimensional context. Our results help to answer questions that arise from analyzing Juno's measurements. 

Until now, an examination of the relation between OCFB and auroral ovals suffered from uncertainties of $>$10° latitude \cite{McGrath2013}. \citeA{Greathouse2022} now present that the auroral ovals, observed by Juno, have a sharp poleward decay and that our modeled surface OCFB matches the bright poleward emission edges in very good agreement. On the downstream side, where the aurora mainly was observed, the latitudinal deviations are $<$1°. Only the Jupiter facing side features little stronger deviations, where the observations are more patchy and our study suggests an increased sensitivity of the OCFB to varied plasma density. A comparison of our model and Juno's observations thus significantly strengthens the conclusion that Ganymede's aurora is brightest exactly at and inside the OCFB.

The various instruments onboard Juno detected the outbound magnetopause crossing more clearly than the inbound, matching expectations of a more dynamic magnetotail without field rotations through the magnetopause. Our model predicts that Juno left Ganymede's magnetosphere at 17:00:16, 14s earlier than JEDI \cite{Clark2022}, 23s earlier than JADE \cite{Allegrini2022} and about 40s earlier than MAG \cite{Romanelli2022} and the Waves instrument \cite{Kurth2022} identified the outbound crossing. Uncertain model parameters could not explain this deviation, leaving an open question for possible further required physics. \citeA{Dorelli2015} for example suggested a thickened double magnetopause induced by the Hall effect at the Jupiter facing side. Except this aspect, our model is in excellent agreement with the Juno MAG and UVS observations.

An entry of Juno into the closed field line region is not consistent with our results. This is also supported by geometrical thoughts as follows. The north-south extent of closed field line region on the downstream side is not expected to increase with distance from the surface. Figures \ref{3dplot} and \ref{auroraplot} reveal that for the closer parts inside the magnetosphere Juno's trajectory, projected with constant $z$ to the surface, was obviously located north of the aurora and correlated surface OCFB and therefore clearly outside the closed field line region.

We assessed model uncertainties through a sensitivity study to uncertain upstream conditions and model parameters, to our knowledge the first of Ganymede's magnetosphere. Our conclusions are robust to these uncertainties and we provide margins for the quantitative results. We found that the variations of all upstream parameters within expected ranges significantly affect different aspects of the magnetosphere and no parameter stands out in its importance. This is also important for the interpretation of the upcoming orbital JUICE or remote-sensing observations without joint in-situ measurements of upstream conditions.

\nocite{Dorelli2015}
\nocite{Fatemi2016}
\nocite{Jia2008}
\nocite{Jia2009}
\nocite{Kivelson2004}
\nocite{Mauk2004}
\nocite{Paty2004}
\nocite{Toth2016}
\nocite{Wang2018}
\nocite{Marconi2007}
\nocite{Roth2021}
\nocite{Ebert2022}
\nocite{Connerney2017a}
\nocite{Duling2022a}
\nocite{Duling2022b}
\section*{Open Research} 
The MHD simulation codes utilized for this work are open-source projects. PLUTO can be downloaded at http://plutocode.ph.unito.it/ (version 4.4). ZEUS-MP is available at http://www.netpurgatory.com/zeusmp.html (version 2.1.2). Juno MAG data are publicly available through the Planetary Data System (https://pds-ppi.igpp.ucla.edu/) at https://doi.org/10.17189/1519711 \cite{Connerney2017a}. The OCFB and Juno's footprint locations on Ganymede's surface data calculated in this study are available at a Zenodo repository via https://doi.org/10.5281/zenodo.7096938 with CCA 4.0 licence \cite{Duling2022a}. The complete simulation output data of our default model are available at a Zenodo repository via https://doi.org/10.5281/zenodo.7105334 with CCA 4.0 licence \cite{Duling2022b}. 

\acknowledgments
This project has received funding from the European Research Council (ERC) under the European Union’s Horizon 2020 research and innovation programme (Grant agreement No. 884711). The research at the University of Iowa is supported by NASA through Contract 699041X with Southwest Research Institute. The numerical simulations have been performed on the CHEOPS Cluster of the University of Cologne, Germany. 


%
%

\bibliography{ExoOceans}

%
%
%
%
%


\end{document}


\nolinenumbers
%
%


\title{Supporting Information for "Ganymede MHD Model: Magnetospheric Context for Juno's PJ34 Flyby"}
%
%

%
%



\authors{Stefan Duling\affil{1}, Joachim Saur\affil{1}, George Clark\affil{2}, Frederic Allegrini\affil{3}, Thomas Greathouse\affil{3}, Randy Gladstone\affil{3,4}, William Kurth\affil{5}, John E. P. Connerney\affil{6,7}, Fran Bagenal\affil{8}, Ali H. Sulaiman\affil{9}}

\affiliation{1}{Institute of Geophysics and Meteorology, University of Cologne, Cologne, Germany}
\affiliation{2}{The Johns Hopkins University Applied Physics Laboratory, Laurel, Maryland, USA}
\affiliation{3}{Southwest Research Institute, San Antonio, Texas, USA}
\affiliation{4}{University of Texas at San Antonio, San Antonio, Texas, USA}
\affiliation{5}{Department of Physics and Astronomy, University of Iowa, Iowa City, Iowa, USA}
\affiliation{6}{NASA Goddard Space Flight Center, Greenbelt, Maryland, USA}
\affiliation{7}{Space Research Corporation, Annapolis, Maryland, USA}
\affiliation{8}{Laboratory for Atmospheric and Space Physics, University of Colorado, Boulder, Colorado, USA}
\affiliation{9}{School of Physics and Astronomy, Minnesota Institute for Astrophysics, University of Minnesota, Minneapolis, Minnesota, USA}

%
%

%


%
%

\noindent\textbf{Contents of this file}
\begin{enumerate}
\item Text S1 to S4
\item Figures S1 to S4
\item Table S1
\item Caption for Movie S1
\end{enumerate}
\noindent\textbf{Additional Supporting Information (Files uploaded separately)}
\begin{enumerate}
\item Movie S1
\end{enumerate}

\noindent\textbf{Introduction}

In S1 we provide a detailed description of the magnetohydrodynamic (MHD) model that was used to obtain the results presented in the article. In S2 we discuss the upstream conditions and other model parameters and their uncertainties. In S3 we describe details of the simulation code PLUTO and the numerical setup used in this analysis. In S4 we shortly compare the results from our independent simulation codes PLUTO and ZEUS-MP, solving the identical physical problem and discuss the impact of the spatial resolution. This discussion is complemented by Figures S1, S2 and Table S1. Figures S3 and S4 show results of all modeled variables on planes that are nearly parallel to Juno's trajectory. Movie S1 visualizes the modeled three-dimensional context of Juno's trajectory during its PJ34 flyby.
\newpage

\noindent\textbf{S1. Model Description}

\label{chmodeldescription}

Ganymede's magnetosphere has been modeled with different physical approaches and numerical solvers. While multi-fluid \cite<e.g.>{Paty2004,Wang2018}, particle-in-cell \cite{Toth2016} and hybrid models \cite{Fatemi2016,Romanelli2022} are primarily excellent for analyzing individual magnetospheric aspects or particle related physics, single fluid magnetohydrodynamic (MHD) \cite<e.g.>{Jia2008,Duling2014} and Hall-MHD \cite<e.g.>{Dorelli2015} models generally allow a higher spatial resolution for numerical reasons and therefore are well suited to model the global interaction topology.

We describe Ganymede's space environment by adopting a MHD model. Since the upstream conditions in Jupiter's magnetosphere can be assumed constant during the time scales of the local interaction at Ganymede the model approaches a steady-state solution. In our single-fluid approach the plasma interaction is described by the plasma mass density $\rho$, plasma bulk velocity $\mathbf{v}$, total thermal pressure $p$ and the magnetic field $\mathbf{B}$. For these variables the MHD equations read in their conservational form, complemented by source terms on their right-hand sides:

\small
\begin{eqnarray}
\frac{\partial\rho}{\partial t} + \nabla \cdot \Big[\rho\mathbf{v}\Big] &=&  P m_n - L m_L,\\
\frac{\partial\rho\mathbf{v}}{\partial t} + \nabla \cdot \Big[\rho\mathbf{v}\mathbf{v} - \frac{1}{\mu_0}\mathbf{B}\mathbf{B} + \mathbf{I} ( p + \frac{1}{2}\frac{B^2}{\mu_0} )\Big] &=& - (L m_L + \nu_n \rho) \mathbf{v},\\
\frac{\partial E_t}{\partial t} + \nabla \cdot \Big[(E_t + p + \frac{1}{2}\frac{B^2}{\mu_0})\mathbf{v}
-  \frac{1}{\mu_0}\mathbf{B}(\mathbf{v}\cdot\mathbf{B}) \Big] &=& - \frac{1}{2} (L m_L + \nu_n \rho) v^2 - \frac{3}{2} (L m_L + \nu_n\rho)\frac{p}{\rho}  + \frac{3}{2}( Pm_n +\nu_n \rho) \frac{k_B T_n}{m_n},\\
\frac{\partial\mathbf{B}}{\partial t} - \nabla \times \Big[\mathbf{v} \times \mathbf{B}\Big] &=& 0 .
\label{eqinduction}
\end{eqnarray}
\normalsize

The total energy $E_t=\frac{1}{2}\rho v^2 + \frac{3}{2}p + \frac{1}{2} \frac{B^2}{\mu_0}$ is composed of the kinetic, thermal and magnetic energy. The model features approximations of physical processes that build on the model of \citeA{Duling2014}. Momentum loss due to particle collisions with neutral O$_2$ molecules is characterized by a collision frequency $\nu_{n}$ as a function of a radially symmetric atmospheric particle density $n_n$. For expected plasma velocities $v_0=140$ km s$^{-1}$ we adopt a constant cross section $\sigma_n=2.2\times10^{-19}$ m$^2$:

\begin{equation}
    \nu_n(r) = \sigma_n v_0 n_n(r) .
\end{equation}

The atmosphere is approximated with a hydrostatic model using a surface density of $n_{n,0}=8.0\times10^{12}$ m$^{-3}$, a constant scale height of $H=250$ km and Ganymede's radius $R_G=2631$ km:

\begin{equation}
    n_n(r) = n_{n,0} \exp{\Bigg(\frac{R_G - r}{H}\Bigg)} .
\end{equation}

Ionization of the atmospheric particles as well as recombination in areas of high density are characterized by the production rate $P$ and loss rate $L$ respectively. We roughly approximate the photo-ionization and electron impact ionization processes by a radially symmetric production rate that is a function of an estimated ionization frequency $\nu_{ion}=2.2\times10^{-8}$ s$^{-1}$:

\begin{equation}
    P(r) = \nu_{ion} n_n(r) .
\end{equation}

The dissociative recombination is parameterized by a recombination rate coefficient $\alpha=7.8\times10^{-14}$ m$^3$ s$^{-1}$ and only active in regions with higher plasma density than the upstream value $\rho_0$:

\begin{equation}
    L = \begin{cases}
    \alpha \rho(\rho-\rho_0)m_L^{-2} & \text{for } \rho > \rho_0 \\
    0 & \text{else}
    \end{cases}
\end{equation}

These parameterizations are explained in detail in \citeA{Duling2014}. For all chemical processes we assume the mass of O$_2$ molecules $m_n=m_L=32$ amu, neglecting the recently detected H$_2$O component on the sub-solar side \cite{Roth2021}. The last term in the energy equation (3) considers the transfer of thermal energy from the neutral atmosphere to the plasma. Since the thermal energy of the atmosphere is low compared to the plasma, this term is expected to be negligible. We keep it for completeness and set the atmosphere's temperature to $T_n=100$ K \cite{Marconi2007} while $k_B$ is the Boltzmann constant.

Ganymede's intrinsic magnetic field is described by dipole Gauss coefficients $g_1^0=-716.8$ nT, $g_1^1=49.3$ nT, $h_1^1=22.2$ nT as derived by \citeA{Kivelson2002}. Within their uncertainties, dipole coefficients updated from Juno data by \citeA{Weber2022} have equal values. Quadrupole models either neglect an induction response of an ocean \cite{Saur2015} or do not significantly improve the fit to available data. In our model we include an induced dipole moment in the equatorial plane:

\begin{equation}
    \begin{pmatrix}
    g_{1, \text{ind}}^1 (\lambda_{\text{III}})\\
    h_{1, \text{ind}}^1 (\lambda_{\text{III}})
    \end{pmatrix}
    = 0.5A 
    \begin{pmatrix}
    B_{0,y}(\lambda_{\text{III}}+\Phi) \\
    -B_{0,x}(\lambda_{\text{III}}+\Phi)
    \end{pmatrix}
    \bigg(\frac{R_T}{R_G}\bigg)^3 .
\end{equation}

Depending on the ocean model the induction response is characterized by the factor $A$ and phase $\Phi$ \cite{Duling2014}, System-III longitude $\lambda_{\text{III}}=302$°, top radius of the ocean $R_T=2481$ km and the inducing field 

\begin{equation}
    B_{0,x}(\lambda_{\text{III}}) = -18 \text{nT} \sin(\lambda_{\text{III}}-200\text{°}),
\end{equation}
\begin{equation}
    B_{0,y}(\lambda_{\text{III}}) = -86 \text{nT} \cos(\lambda_{\text{III}}-200\text{°}).
\end{equation}

During Juno's visit Ganymede was near the center of the current sheet where the induction response is close to minimum. The corresponding factor $A=0.95$ and phase $\Phi=-7$° result in a maximum surface strength of the induced dipole of 15.6 nT.

The occurrence of magnetic reconnection has been confirmed at Ganymede \cite{Ebert2022,Romanelli2022}. To adequately model reconnection the consideration of finite plasma conductivity at the magnetopause is necessary. Theoretically, our model allows the addition of a physical resistivity term to the induction equation \ref{eqinduction}. In a simulation with lower grid resolution (50-250 km cell size inside the magnetosphere) we included anomalous and ionospheric resistivity similar to \citeA{Duling2014} and \citeA{Jia2009} and have not found any significant impact on the magnetic topology at all. The strongest deviations from the same simulation without physical resistivity were $<$1 nT for the magnetic field on Juno's trajectory and $<$0.1° for the OCFB surface location. To reduce the computing time of simulations with high grid resolution (S3) and
make extensive parameter studies possible we therefore deactivated the resistivity term for the presented study. The always present numerical resistivity, that results from the discretization of space and time and depends on the solver, reduces with higher grid resolution. However, by increasing the resolution we found that the solution globally converges to better fit the measurements, suggesting better results with reduced resistivity (Section S4). Therefore we still expect the impact of the physical resistivity on the high resolution simulations would be low and insignificant for the results of this study.

\noindent\textbf{S2. Upstream Conditions and Model Parameters}

Our model uses homogeneous and steady-state upstream conditions that are adjusted to the situation during Juno's flyby. Most of the values are not available yet from direct measurements. Therefore we use empirical or modeled predictions which have uncertainties of different order. To assess the model sensitivity on these parameter uncertainties (Section 4) we consider the minimum and maximum realistic values.

An appropriate value for Jupiter's magnetospheric field at the location of Ganymede can directly be obtained from measurements of the magnetometer on-board Juno. Therefore the undisturbed field measurements before (~(-16,3,-70) nT) and after (~(-14,43,-80) nT) the flyby \cite{Weber2022} have to be interpolated to obtain a value suitable for the situation during CA (~(-15,24,-75) nT). This value has some uncertainties because the temporal change is possibly non-linear and the convection time might play a role as well. The measurements before and after CA are nevertheless most likely upper and lower limits for the magnetic field.

Upstream plasma conditions are more difficult to determine. Juno's JADE and JEDI instruments provide particle distribution functions which in theory enable numerical moment calculations to achieve the plasma density, velocity and thermal pressure. However, at the moment numerical moments do not provide reliable values. Until refined analysis might help to determine those upstream conditions in the future, we access predictions. The plasma velocity relative to Ganymede depends on how strongly Jupiter's magnetosphere sub-corotated during the flyby. Voyager and Galileo data suggest a relative velocity of 140 km/s with a variability of 20 km/s \cite{Kivelson2022}. The density is expected to vary by a factor of 5 depending on Ganymede's position with respect to the current sheet \cite{Jia2008}, whereas literature values reveal larger uncertainties: 54 amu/cm$^3$ on average with a variability of 2-100 amu/cm$^3$ \cite{Kivelson2004}, 30 (13-46) amu/cm$^3$ with an uncertainty factor of 2 \cite{Bagenal2011}, 160 amu/cm$^3$ inside the current sheet and 48 amu/cm$^3$ on higher magnetic latitudes \cite{Kivelson2022}. JADE measured ~1 cm$^{-3}$ protons and ~8 cm$^{-3}$ heavy ions before the flyby \cite{Allegrini2022}, consistent with electron densities of 5-12 cm$^{-3}$ observed by the Waves instrument outside of the magnetosphere \cite{Kurth2022}. We assume 100 amu/cm$^3$ for our model and investigate the effects of extreme densities 10 and 160 amu/cm$^3$. The thermal pressure is dominated by energetic particles in the vicinity of Ganymede \cite{Mauk2004}. Therefore JEDI measurements provide a lower limit to the pressure during the flyby. \citeA{Clark2022} calculated 1.5 nPa for the $>$50 keV protons. Sulfur and oxygen are expected to have a significant but unknown contribution. Former models assumed 3.8 nPa \cite{Jia2008,Duling2014}, here we use 2.8 nPa as also specified by \citeA{Kivelson2022} and consider generous limits of 1.0 nPa and 5.0 nPa as uncertainties.

The best guess upstream conditions for our default model characterize the interaction to be sub-Alfv\'enic with an Alfv\'en Mach number of $M_A=$0.8,  and a plasma beta of 1.1. 

\noindent\textbf{S3. Numerical Solution Process}

\label{chmodelnumerics}
We perform numerical simulations to obtain an approximate solution for equations (1-4). While we utilized the ZEUS-MP code \cite{Hayes2006} for our former work, we now present results obtained with the PLUTO code \cite{Mignone2007}. This code is broadly used in the plasma science community and gives us the advantage to compare and validate our model results obtained by two different and independent numerical solvers. PLUTO is an open-source software designed to solve hyperbolic and parabolic systems of PDE's for astrophysical fluid dynamics. In contrast to ZEUS-MP's finite difference approach it uses the finite volume method. In our application we utilize a piece-wise linear, 2nd order reconstruction of the variables in the cells, the Harten, Lax, Van Leer solver for the Riemann problem to calculate the fluxes at the cell interfaces and a 2nd order Runge Kutta scheme for the integration forward in time. Unfeasible time steps and instabilities caused by possibly emerging vacuums are prevented by ensuring a minimal mass density and thermal pressure of 5\% of the upstream value. 

For the numerical solution we divide the space between Ganymede's surface and 70 Ganymede radii ($R_G$) into a grid with spherical geometry and a longitudinal resolution of 1.4°. To adequately resolve the strong magnetic tension in the equatorial region we use a latitudinal resolution of 0.74° between 26°N/S and 1.4° at the poles. Below 1.2 $R_G$ the radial resolution equals 0.017 $R_G$ and increases afterwards smoothly in the region below 12 $R_G$ and steeply afterwards. Near the magnetopause at $\sim$ 2 $R_G$ the radial cell distance is 0.025 $R_G$, at 12 $R_G$ it is 0.12 $R_G$ and at the outer boundary at 70 $R_G$ it is 1.7 $R_G$. With a radial and latitudinal resolution of $<$65km inside the closed field line region the grid is able to resolve the ion inertial length ($\sim$320km) \cite{Dorelli2015}. The cell number resolution is 308$\times$208$\times$256 $(r, \theta, \phi)$ resulting in 16.4 million cells in total.

Representing Ganymede's surface, the inner boundary absorbs the incoming plasma. This is considered by applying open conditions for the plasma variables in addition to forcing the radial velocity component to be zero or negative. Ganymede's icy, electrically non-conducting crust cannot carry electric currents. This property directly affects the near surface magnetic field and is considered through isolating boundary conditions derived in \citeA{Duling2014}. At the outer boundary we use fixed boundary values equal to the upstream conditions on the upstream side and open conditions on the downstream side.

\noindent\textbf{S4. Impact of Solver and Resolution on the Results}

\label{chzeuscomparison}
In addition to the PLUTO code \cite{Mignone2007}, applied in this work, we previously modeled Ganymede with the ZEUS-MP code \cite{Hayes2006}. This gives us the unique opportunity to compare the results of identical models setups numerically calculated with two independent solvers. However, using the high spatial resolution of this study with ZEUS-MP exceeds the technical resources we have available. Therefore we performed lower resolution simulations with both codes for the identical model described in Section S1 and our default parameter set. In this comparison we use 2.1 million cells in total, the minimal radial resolution is 0.02 $R_G$ and the constant latitudinal and longitudinal resolutions are 2.8°. Within the closed field line region the radial and latitudinal resolution is $<$260km.

\begin{table}
\caption{Results for the identical physical model obtained from independent simulation codes for different spatial resolutions. Columns 3-6 specify the averaged latitude of the northern and southern open closed field line boundary (OCFB) on Ganymede's surface on the upstream (-45° to -135°W) and downstream (45° to 135°W) side. Column 7 lists Juno's closest distance to closed field lines (CF) and columns 8-9 the UTC times of its inbound and outbound magnetopause crossings, respectively. Column 10 lists the RMS between measured and modeled magnetic field between 16:50 and 16:59.}
    \centering
    \begin{tabular}{ l l l l l l l l l r}
        & & \multicolumn{2}{c}{OCFB down}& \multicolumn{2}{c}{OCFB up} & CF & \multicolumn{2}{c}{magnetopause crossing} & RMS\\
        code & resolution & N [°] & S [°] & N [°] & S [°] & [$R_G$] & inbound & outbound & [nT] \\ 
        \hline
        PLUTO$^a$        & high        & 21.2 & -24.4 &  51.5 & -47.4 & 0.26 & 16:48:16 & 17:00:16 &  9.3\\
        PLUTO            & low         & 27.6 & -30.8 &  54.3 & -50.3 & 0.13 & 16:49:25 & 17:00:26 & 23.5\\
        ZEUS-MP             & low         & 28.1 & -31.2 &  56.4 & -51.8 & 0.02 & 16:48:50 & 17:00:00 & 26.9\\
        \multicolumn9l{$^a$: default model of the main study}\\
    \end{tabular}
\end{table}

\begin{figure}
\centering
\noindent\includegraphics[width=0.8\textwidth]{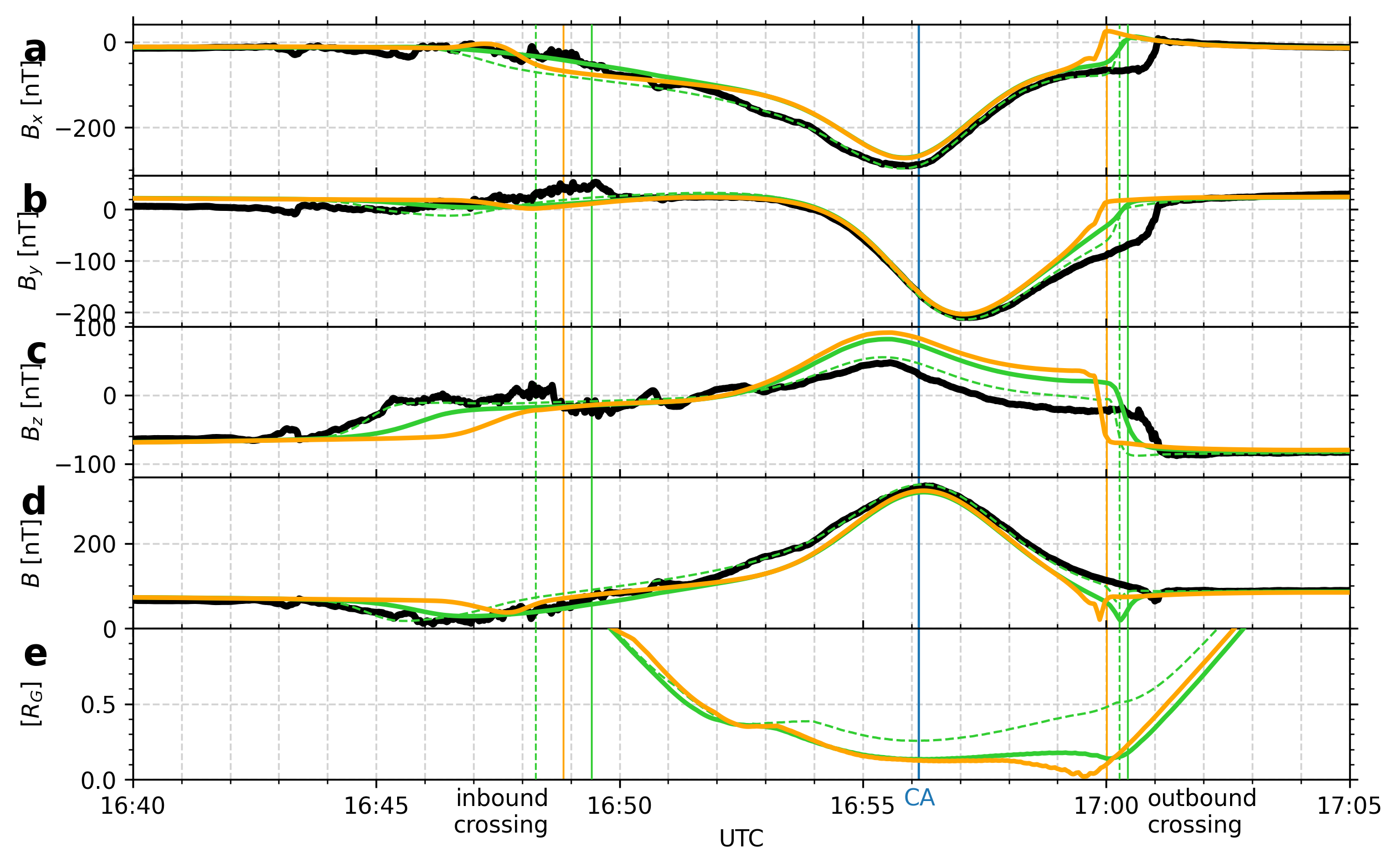}
\caption{Modeled magnetic field along Juno's trajectory from PLUTO (green) and ZEUS-MP (orange) for the identical physical model and reduced spatial resolution. The magnetic field from the default model with higher resolution of the main study is shown as dashed green lines, Juno's measurements in black. Panels a-c show GPhiO components, panel d the magnitude. Panel e shows Juno's distance from the OCFB in $R_G$. The vertical lines represent the modeled inbound and outbound magnetopause crossings.}
\label{Strajplot}
\end{figure}

\begin{figure}
\centering
\noindent\includegraphics[width=0.8\textwidth]{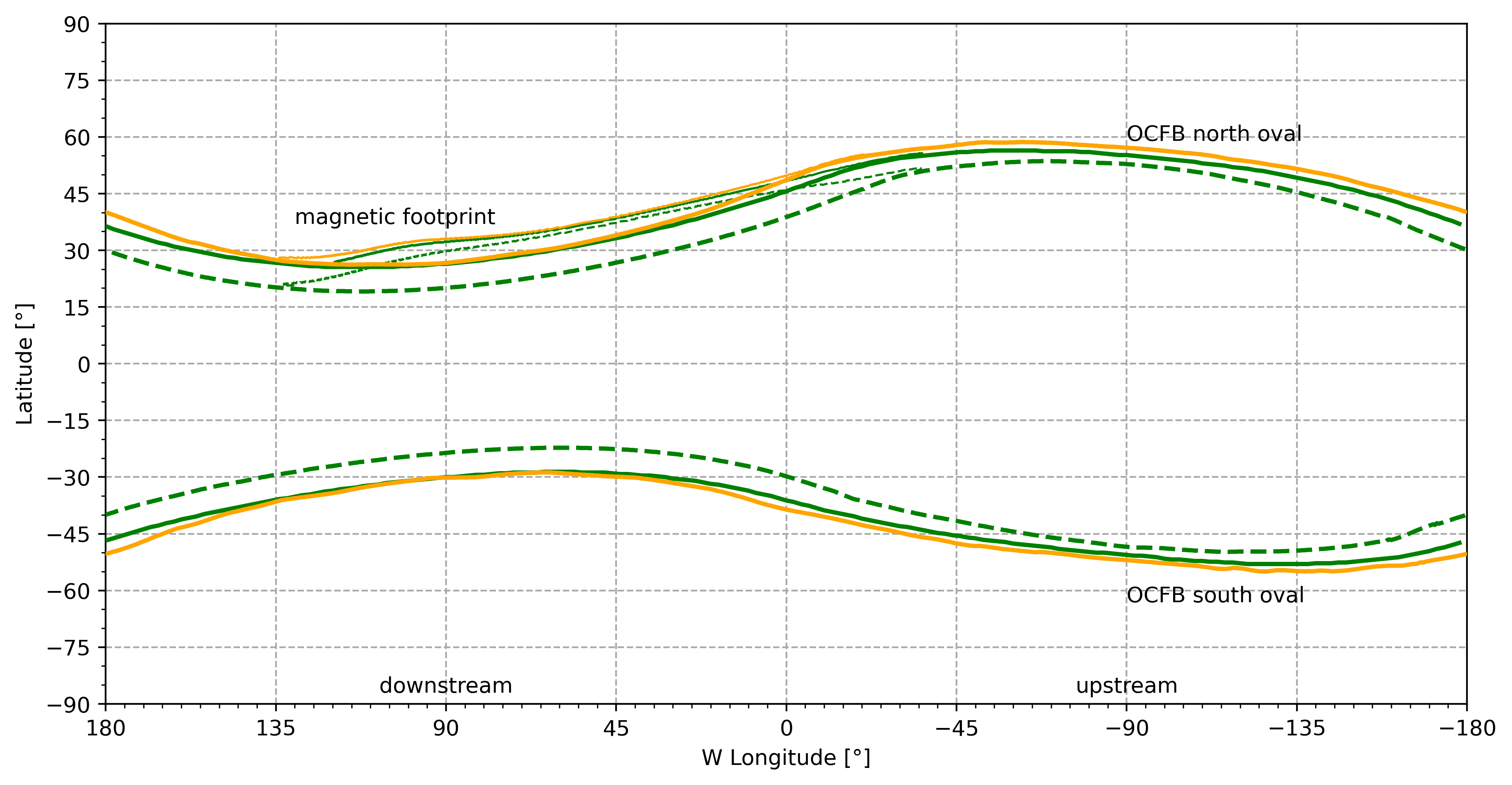}
\caption{Surface map of Ganymede with 0° western longitude pointing towards Jupiter ($+y$ axis GPhiO). The OCFB (thick) and Juno's magnetic footprint (thin) from PLUTO (green) and ZEUS-MP (orange) but the identical physical model are shown for a reduced spatial resolution. The results from the default model of the main study are shown as dashed green lines.}
\label{Sauroraplot}
\end{figure}

In Figures \ref{Strajplot} and \ref{Sauroraplot}, similar to Figures 3 and 4 of the main study, we present the results of PLUTO (green) and ZEUS-MP (orange) together with the PLUTO results with the higher resolution of the main study (dashed green). Table S1 lists selected quantitative results. First, we shortly analyze the effect of a reduced spatial resolution by considering only the results from PLUTO. In general the modeled magnetic field fits the observations worse than with the high resolution, especially the $B_z$ component (Figure \ref{Strajplot}c) at the inbound field rotation (16:45) and near the closed field line region around CA. Additionally the OCFB on Ganymede's surface is shifted polewards (Figure \ref{Sauroraplot}) for all longitudes, indicating a larger closed field line region. As consequence the closest distance to closed field lines (Figure \ref{Strajplot}e) is reduced, but Juno is still not entering the closed field line region. An effect of a reduced spatial resolution is the increase of the numerical resistivity. The code becomes more diffusive and as consequence boundary layers become less sharp pronounced, magnetic tension is less preserved and the closed field line region expands. Since the high resolution results fit the observations very well we expect a weaker resistivity to improve the fit and the high resolution to be sufficient to model the Juno flyby. However, the location of the outbound magnetopause crossing (Figure \ref{Strajplot}, $\sim$17:00:26) is better modeled with the lower resolution, even though the shape of the magnetic field rotation is worse. As discussed in the main study the outbound magnetopause crossing remains an open question.

Comparing the results from PLUTO and ZEUS-MP for the low resolution we find similar results. The OCFB on Ganymede's surface (\ref{Sauroraplot}) has differences between 5° at the anti-Jovian side and $<$1° at the downstream side where Juno mainly observed the aurora. The OCFB from ZEUS-MP is located polewards of the OCFB from PLUTO which is also reflected in Juno's slightly lower distance to closed field lines in the case of ZEUS-MP (Figure \ref{Strajplot}e). The modeled magnetic field from ZEUS-MP (Figure \ref{Strajplot}c) also has a slightly worse fit to the field rotation in the wake region (around 16:45) and the outbound magnetopause crossing. All these discrepancies suggest that ZEUS-MP might be slightly more resistive than PLUTO. For the major parts of Juno's trajectory, however, both independent codes produce very similar results, suggesting additional reliability of our numerical implementations. The primary purpose of this comparison is that the remaining differences give an impression of the numerical error that can be expected while modeling Ganymede's magnetosphere.

\noindent\textbf{Movie S1}

The movie illustrates the three-dimensional geometry of Ganymede's magnetosphere in reference to Juno's trajectory (red line) for the time around closest approach. The green surface represents the extent of the closed field line region. The blue surfaces represent the regions with open field lines that connect Ganymede's polar regions with Jupiter and correspond to the Alfv\'en wings. The white tubes show selected closed and open field lines and the orange tubes show field lines that are seeded on Juno's trajectory. Outside of the magnetosphere these field lines are unconnected and inside the magnetosphere they end at Ganymede's surface, representing Juno's magnetic footprint. Auroral oxygen emissions are displayed on Ganymede's surface as observed by Juno's UVS instrument \cite{Greathouse2022}.

\begin{figure}
\noindent\includegraphics[width=\textwidth]{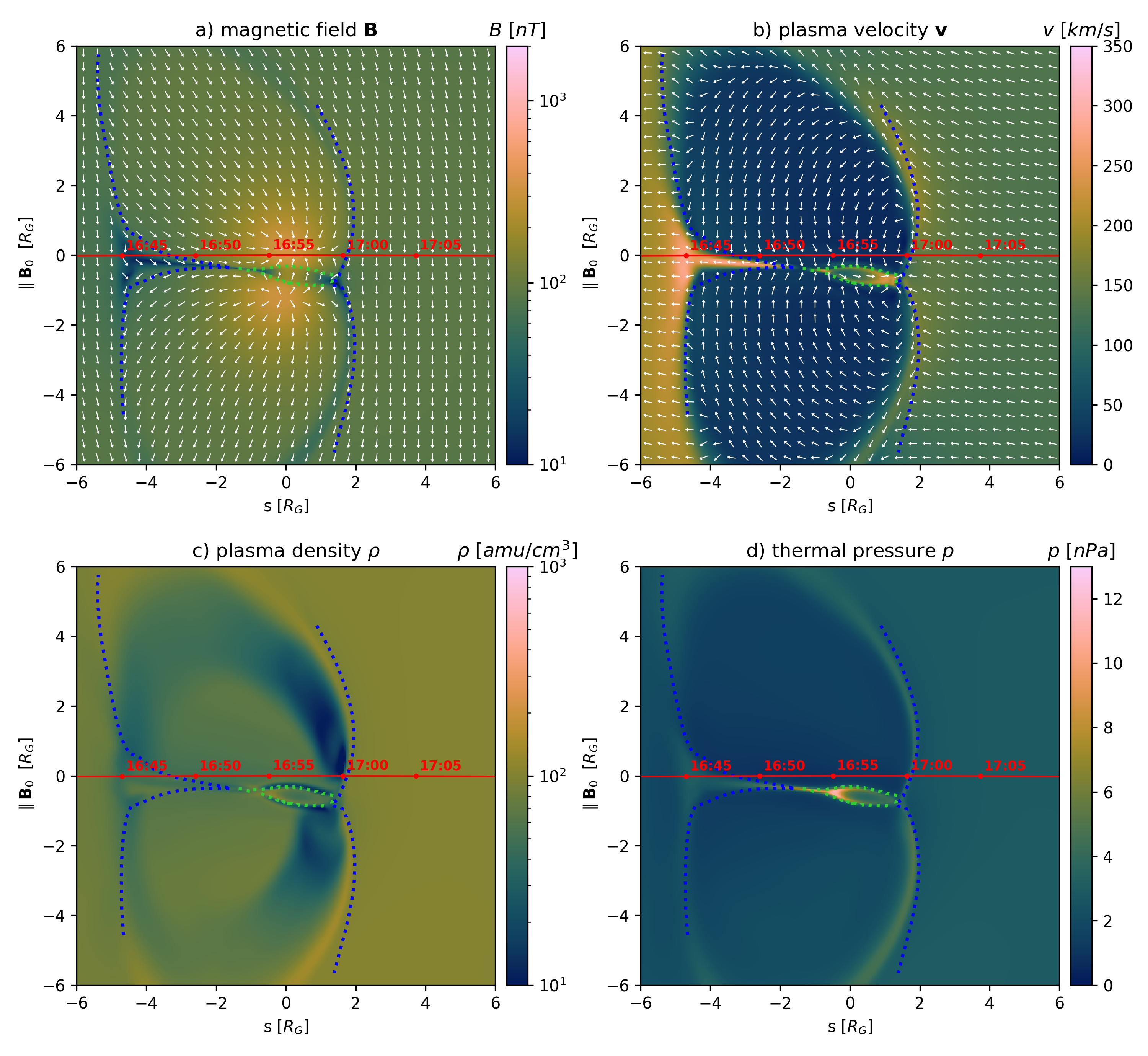}
\caption{Model variables for Juno's flyby on a plane that is spanned by Juno's trajectory and the direction of the upstream magnetic field. The origin of the plane is the location of Juno's closest approach to Ganymede. The $s$-coordinate is along the velocity vector of Juno at closest approach, the second direction is as parallel to the upstream magnetic field as possible. The red line shows Juno's trajectory that projected is only minimally. The green dotted line represents the intersection of the OCFB with the plane, the blue dotted lines represent the intersection with the magnetopause. The white arrows show the projected direction of $\mathbf{B}$ and $\mathbf{v}$ respectively.}
\end{figure}

\begin{figure}
\noindent\includegraphics[width=\textwidth]{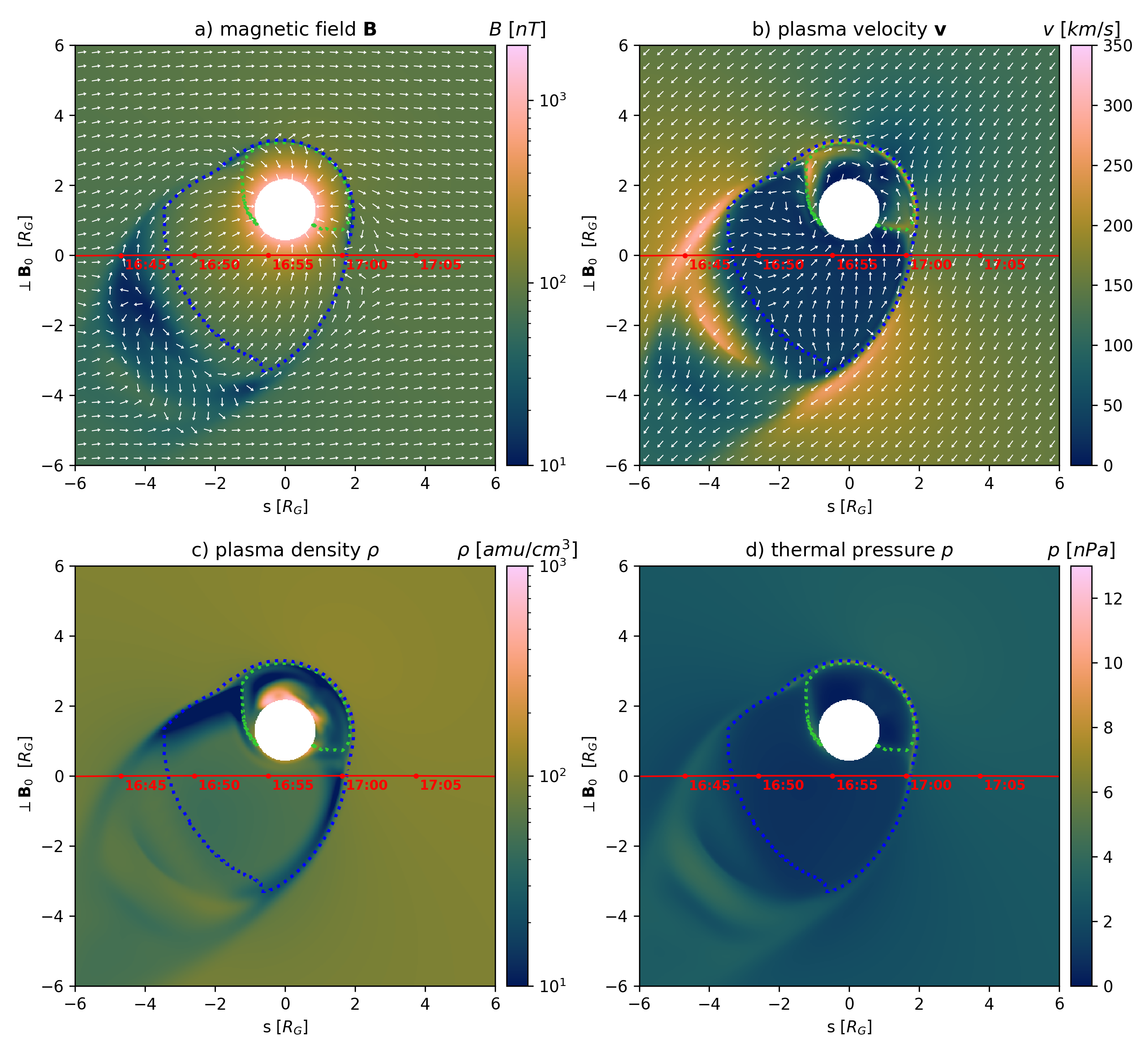}
\caption{Model variables for Juno's flyby on a plane that is spanned by Juno's trajectory and the direction perpendicular to the upstream magnetic field. The origin of the plane is the location of Juno's closest approach to Ganymede. The $s$-coordinate is along the velocity vector of Juno at closest approach, the second direction is as perpendicular to the upstream magnetic field as possible. The red line shows Juno's trajectory that is projected only minimally. The green dotted line represents the intersection of the OCFB with the plane, the blue dotted line represents the intersection with the magnetopause. The white arrows show the projected direction of $\mathbf{B}$ and $\mathbf{v}$ respectively.}
\end{figure}

\newpage


%
%


%
%
%
%
%

\bibliography{ExoOceans}